Title: Single-cell measurement of red blood cell oxygen affinity

Authors: Giuseppe Di Caprio[a,1], Chris Stokes[a], John M. Higgins[b,c], and Ethan Schonbrun[a]

[a] The Rowland Institute, Harvard University, Cambridge, MA 02142
[b] Center for Systems Biology, Massachusetts General Hospital, Boston, MA 02115
[d] Department of Systems Biology, Harvard Medical School, MA 02115
[1] corresponding author: Giuseppe Di Caprio dicaprio@rowland.harvard.edu, tel +1 (617) 497-4714

Keywords: hematology, medical sciences, flow cytometry, organ-on-chip, single-cell variability

Short title: Variability in red blood cell oxygen affinity

Abstract

Oxygen is transported throughout the body by hemoglobin in red blood cells. While the oxygen affinity of blood is well understood and is routinely assessed in patients by pulse oximetry, variability at the single-cell level has not been previously measured. In contrast, single-cell measurements of red blood cell volume and hemoglobin concentration are taken millions of times per day by clinical hematology analyzers and are important factors in determining the health of the hematologic system. To better understand the variability and determinants of oxygen affinity on a cellular level, we have developed a system that quantifies the oxygen saturation, cell volume and hemoglobin concentration for individual red blood cells in high-throughput. We find that the variability in single-cell saturation peaks at an oxygen partial pressure of 2.5%, which corresponds to the maximum slope of the oxygen-hemoglobin dissociation curve. In addition, single-cell oxygen affinity is positively correlated with hemoglobin concentration, but independent of osmolarity, which suggests variation in the hemoglobin to 2-3 DPG ratio on a cellular level. By quantifying the functional behavior of a cellular population, our system adds a new dimension to blood cell analysis and other measurements of single-cell variability.

Significance

Oxygen transport is the most important function of red blood cells. We describe a microfluidic single-cell assay that quantifies the oxygen saturation of individual red blood cells in high throughput. While single red blood cell measurements of volume and mass are routinely performed in hospitals by hematology analyzers, measurements that characterize the primary function of red blood cells, the delivery of oxygen, have not been made. With this system, we find measurable variation that is positively correlated with cellular hemoglobin concentration, but independent of osmolarity. These results imply that the cytoplasmic environment of each cell is different and that these differences modulate the function that each cell performs.

Introduction

Red blood cells are the most common type of blood cell and constitute approximately half of the human body's total cell count [1]. They take up oxygen in the lungs and deliver it throughout the body, taking on average 20 seconds to complete one circuit through the circulation [2]. Each cell is densely packaged with hemoglobin that binds and releases oxygen based on the local oxygen partial pressure. The fraction of occupied binding sites relative to the total number of binding sites is called the oxygen saturation and can be described by the hemoglobin-oxygen dissociation curve. While it is known that several factors affect the oxygen affinity of hemoglobin and consequently shift the dissociation curve, such as pH, temperature and 2-3DPG [3], it is not known how much variation these factors cause on a cellular level within individuals.

Recently there has been significant progress in developing biomimetic environments for studying physiological processes of cells *in vitro*. Analogs to the lung [4], heart [5], bone marrow [6], and gut [7] have all been developed in microfabricated chips that enable *ex vivo* studies to closely replicate an *in vivo* environment. In addition to adherent cell cultures, chips for studying flowing blood have been developed that enable control of oxygen partial pressure with high spatial resolution [8-12]. We take advantage of this new technology and combine it with a recently developed microfluidic cytometry method that enables us to quantify cell volume and hemoglobin mass for individual flowing cells in high-throughput [13].

Our system for measuring red blood cell mass differs from previous single-cell mass measurements [14-18] in that it is based on the optical absorption of hemoglobin. Unlike previous cell mass measurements, however, it is straightforward to extend this method to resolving the mass of both oxygenated and de-oxygenated species because of the well-known modification of hemoglobin's absorption spectra due to oxygen binding [19]. This allows us to quantify red blood cell volume, hemoglobin concentration and oxygen affinity for cells while they are in a fluidic environment similar to the circulatory system. A few previous studies have explored the measurement of single-cell saturation [20-22], but to our knowledge, this has never been performed on a large cell population or under accurate control of oxygen partial pressure.

In this paper, we first describe the physical properties of the microfluidic chip, followed by the optical measurement system for obtaining cell volume and hemoglobin mass. We then discuss the spectroscopic measurement system and the quantification of single-cell saturation from the measured multispectral absorption. The temporal dynamics of the system are then characterized to better understand the time and length scale required for equilibrium de-oxygenation. We use our system to capture a hemoglobin-oxygen dissociation curve and a standard deviation curve as a function of oxygen partial pressure. At each oxygen partial pressure, we can retrieve a full distribution of single cell oxygen saturation and observe its correlation to total hemoglobin concentration. In addition, using the measured saturation values, we can retrieve a $P_{50}$ value, the oxygen partial pressure where the saturation is 50%, for each cell in the population using the Hill equation as a model.

A functional red blood cell analyzer

The microfluidic device to control the oxygen partial pressure and to deliver cells to the measurement region is shown in Fig. 1 a, b. The device is composed of three layers, one level in which the RBCs flow, a thin gas-permeable membrane (PDMS, thickness=90 μm) and the gas channel (section shown in Fig 1b). Cells are driven through the microfluidic device using a syringe pump and flow through a gas exchange region that is 1 cm long, 1 mm wide and 6 μm thick. As they flow, cells equilibrate with the oxygen partial pressure in the gas channel located above a gas-permeable membrane. A hexagonal lattice of PDMS pillars (50 μm of diameter, separated by 100 μm) sustains the channel (left inset in Fig. 1a). Traveling at a mean velocity of 2 mm/s, cells spend approximately 5 seconds in the gas exchange region, which is comparable to the time that they spend in the microcirculation [23]. The gas channel is serpentine and has dimensions of 250 μm wide and 35 μm thick. The oxygen partial pressure of the input gas is controlled off-chip by mixing a tank of pure N2 with a tank of air (21% O2 and 79% N2). The partial pressure of the gas mixture is measured by an oxygen sensor (*GS-Yuasa Oxygen Sensors KE-Series*) before and after flowing in the gas serpentine channel in order to verify that the system has reached a steady-state condition. After cells pass through the gas exchange region, they are imaged in the measurement region which is composed of 16 parallel channels (width 30 μm), shown in the right inset of Fig. 1a.

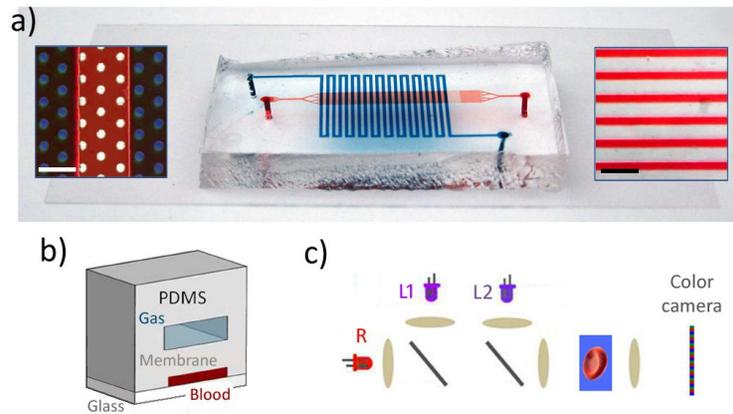

Fig. 1 Gas-exchange red blood cell cytometry. a) Shows a side view of the microfluidic chip. The gas and the sample channel are filled with blue and red dye, respectively. The sample channel is 2.2 cm long. The left inset shows a picture of the gas-exchange region, where the pillar structure is visible as well as the overlay between the gas and the sample channel. The right inset shows the parallel channels in the measurement region. In both insets scalebars are 200 μm. b) Sketch of the cross-section of the three layers composing the chip. c) Shows the optical setup. Two blue (L1 and L2) and one red (R) LEDs are combined using dichroic mirrors. The light transmitted by the cells flowing in the microfluidic channel is then projected onto a color camera by a microscope objective.

The optical setup is shown in Fig. 1c. Cells are suspended in a refractive index-matching buffer into which a non-membrane permeable absorbing dye has been added (AB9, concentration of 0.8 g/dL) [13, 24]. The cells are illuminated using a red LED ($\lambda$ = 625 nm) to quantify cell volume. The presence of a cell in the detection region displaces dye molecules in the optical path and consequently increases the transmitted light signal. If there is minimal absorption or scattered light from the cell, the change in optical transmission can be related to the absolute cell height at each pixel of the image. The refractive index of the buffer has been increased to approach the average cell refractive index by adding 33 g/dL of bovine serum albumin (BSA). To quantify total hemoglobin mass, which is the sum of oxygenated ($HbO_2$) plus deoxygenated hemoglobin (Hb) mass, we use our previously demonstrated method of quantitative absorption cytometry [13]. Oxygenated and de-oxygenated hemoglobin ($HbO_2$ and Hb respectively) have different absorption spectra. Consequently by acquiring images of the same cell at wavelengths corresponding to the two absorption peaks (L1 and L2 in Fig. 1.c), the two species of hemoglobin in the cell can be decoupled and quantified (see *Experimental procedures*). Single-cell saturation is defined as the ratio of $HbO_2$ divided by the total hemoglobin mass.

The kinetics of oxygen binding and release

Because our goal in this study is to measure the oxygen saturation of cells that are at equilibrium with their environment, we need to first characterize the oxygen binding kinetics of flowing red blood cells. For this measurement, we use a slightly different chip geometry, shown in Fig. 2a, where cells flow in straight channels for the entire duration of the gas exchange, which enables the oxygen binding and release kinetics to be studied. By measuring the value of the saturation at different intersections of the blood channel within the gas serpentine channel, we can quantify at what location and time the cells reach a stable oxygen saturation.

The mean cell saturation is shown at 8 different positions for two different gas mixtures ($ppO_2$ = 0 % and 2.95 %), with the mean cell saturation reaching equilibrium at 0 % and 40 % respectively (Fig. 2b). It is evident from the plotted data that cells reach a stable saturation at around the tenth intersection. To further quantify the kinetics, we have fit the data using the decay function: $Sat = Ae^{(-\frac{x}{\tau})} + B$ (Fig. 2b). As a function of the intersection number, we obtain a decay constant of $\tau$ = 3.2. Making use of the average cell velocity, we obtain a deoxygenation decay time of $\tau$ = 800 ms, which is comparable to the time that cells spend in the lung alveoli [23]. For measurements taken beyond the tenth intersection, we have found no correlation between cell saturation and cell velocity or cell position. We consequently believe that in the final measurement region, all cells are at a steady state saturation and experience the same oxygen partial pressure. This results in variation of cell saturation that is caused exclusively by the variants of hemoglobin that are in each cell and by the cell's unique cytosolic environment. Figure 2c shows the $HbO_2$ and Hb mass maps of different cells as a function of the intersection number at 0% $ppO_2$. From these images, it is also clear that cells have a constant oxygen saturation after the tenth intersection.

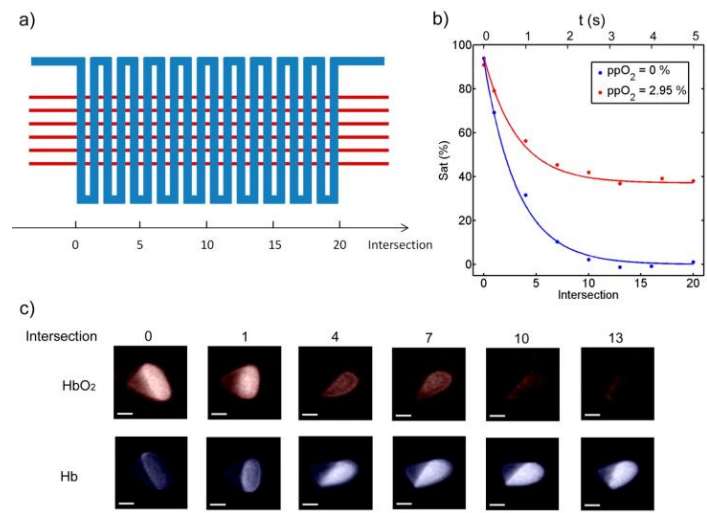

Fig. 2 Red blood cell oxygen binding kinetics. a) Shows a schematic of the chip geometry used to characterize cellular binding kinetics. Blood flows in the red straight channel, from left to right, and the saturation is measured at different intersections with the gas channel (in blue); b) Shows the saturation at different intersections for two different gas mixtures, plotted vs. intersection number and time, for a cell velocity of 2 mm/s. c) Oxygenated ($HbO_2$) and deoxygenated (Hb) hemoglobin mass maps of flowing cells at different intersections. Scalebar is 3 µm.

Variability in cellular oxygen affinity

By scanning ppO$_2$ from 0-21 %, we can evaluate the mean and the standard deviation (σ) of the single-cell saturation distribution, as shown in Fig. 3a, plotted in blue and red respectively. The curve of the mean saturation values is consistent with the expected cooperative binding model for hemoglobin. To fit the mean saturation data, we use the Hill function,

$$Sat = \frac{1}{\left(\frac{K_d}{ppO_2}\right)^n + 1}, \qquad (2)$$

where $K_d$ is a dissociation constant and $n$ is the Hill coefficient. The strength with which oxygen binds to hemoglobin is affected by several factors, such as temperature, pH, Hb variants, and the ratio of 2-3 DPG to hemoglobin. This data, and the rest of the data in this paper unless otherwise stated, was collected at 22°C and a pH of 5.5, due to the acidic properties of BSA. For the curve plotted in Fig. 3a, the Hill coefficient is 3.5 and $K_d$ is also 3.5. Figure 3b shows the measured dissociation curve of the same sample under two other experimental conditions, one at physiological conditions of 37°C and a pH of 7.2, where NaOH has been added to the buffer to neutralize the pH (measured using a *ORION PerpHecT PH meter*), and one at 37°C and a pH of 5.5. The sample at physiological conditions has a similar oxygen affinity, showing a slightly right shifted dissociation curve. The sample at 37°C and low pH shows a substantially lower oxygen affinity, which is expected and caused by the Bohr Effect [2].

    In addition to mean cell values, we can for the first time report σ of the saturation distribution as a function of oxygen partial pressure, which ranges from 1.2 % to 2.9 %, when ppO$_2$ is 8 % and 2.8 %, respectively for this sample. The low σ of 1.2 % at 8% ppO$_2$ is expected because the dissociation curve is almost flat at high pressures. We can consequently use this value, approximately 1%, as a conservative estimate of the single-cell saturation measurement accuracy of our system. At each ppO$_2$ we analyze approximately 2000 cells, so our estimated accuracy of the population mean is $\sigma/\sqrt{n}$ < 0.1% which is similar to the uncertainty in standard deviation for such a large sample size. Conversely, σ peaks when the slope of the dissociation curve is at its maximum and at these pressures the variation in saturation of different cells is at its maximum value. Interestingly, this results in a much larger cellular variation at venous than arterial ppO$_2$ [25]. Across the entire range of pressure, we find that the standard deviation curve is proportional to the slope of the dissociation curve, which we have used as a fit in Fig. 3a.

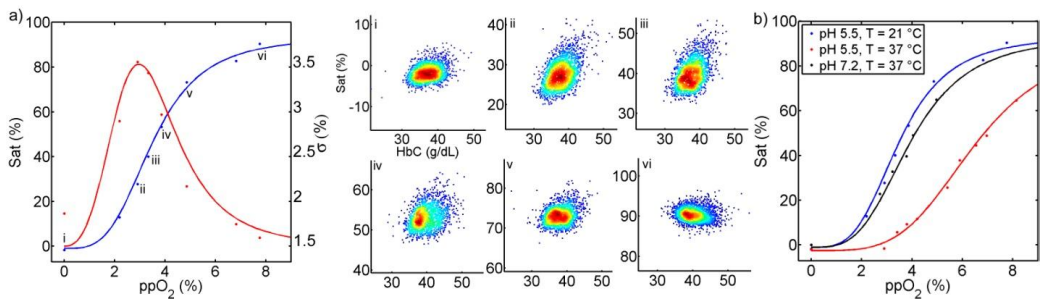

Fig. 3 Single-cell saturation distribution. a) Shows the mean (blue points) and standard deviation (red points) of measured saturation at different values of ppO$_2$. The blue and the red lines plot the Hill

function and its derivative fitting the saturation values and the standard deviation, respectively. i-vi) Scatter plots of saturation vs. HbC for 6 different ppO$_2$ are shown at the right. b) Saturation values and the dissociation curve measured for different pH and temperature. The curves show the Hill function fits performed on the three sets of data.

In addition to the mean and standard deviation, we can plot the entire saturation distribution and investigate its correlation with red blood cell volume and hemoglobin concentration. Figure 3a shows scatter plots of single cell saturation and hemoglobin concentration for the same sample at six different ppO$_2$. As can be seen most easily in Fig. 3a.ii-iii, these two variables show an interesting positive correlation, where cells with high Hb concentration (HbC) show high saturation values. The measurement has been repeated on several different samples from different individuals and each displays a positive correlation between HbC and cell saturation at pressures that correspond to the peak in variability. See *S1* for more details.

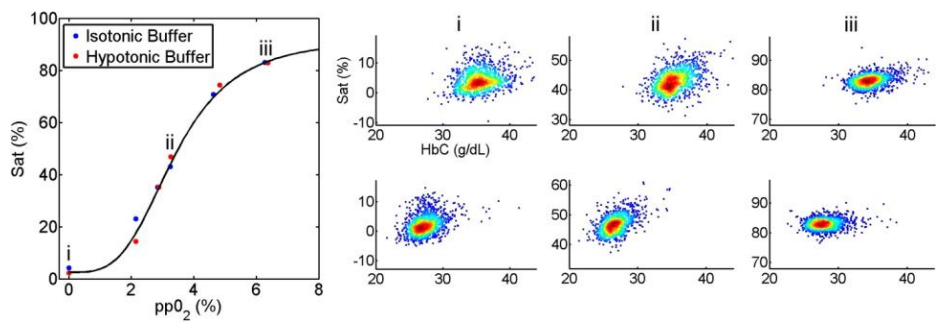

fig. 4 Single-cell saturation and hemoglobin concentration. The plot shows saturation values for cells suspended in an isotonic (272 mOsm) and hypotonic (175 mOsm) buffer. The black line plots the Hill function fitting the saturation values in both regimes. Scatter plots of saturation vs. HbC for 3 different ppO$_2$ are showed for comparison between the two buffers with different osmolarity.

To further investigate the correlation of single-cell saturation with HbC, we perform measurements on the same sample in both an isotonic (272 mOsm) and a hypotonic (175 mOsm) buffer. If single-cell oxygen affinity is solely due to cellular HbC, then one would expect the dissociation curve to shift when the HbC is reduced by the hypotonic buffer. We measured the mean HbC in the isotonic buffer to be 35.6 g/dL, whereas the mean HbC in the hypotonic buffer was reduced to 27.4 g/dL. As can be seen in Fig. 4, however, the same Hill function fits the saturation values in the isotonic and the hypotonic regime. In addition, from the scatter plots, we see that while the distribution of HbC shifts to the left in the hypotonic buffer, the shape and location of the cloud is maintained. It has been proposed that hemoglobin's oxygen affinity is modulated by its extremely high intracellular concentration in a process called molecular crowding, but this would produce a shift caused by the osmotic change in volume [26]. Instead, we suspect that the observed correlation between HbC and cell saturation is due to variation across the sample in the ratio between hemoglobin and 2-3DPG molecules inside each cell [27, 28]. Osmotic perturbation will not change this ratio.

While saturation values determine the oxygen content of a blood sample at a given ppO$_2$, it is often convenient to use a single number to quantify the oxygen affinity across a range of ppO$_2$. For this purpose, clinicians frequently use a blood sample's $P_{50}$. Using the Hill equation as a model, our measurement allows the evaluation of the $P_{50}$ at a single-cell level. In Fig. 5a we plot a histogram of the saturation distribution at $PO_2$ = 2.92 % and the resulting $P_{50}$ distribution. Figure 5b shows the derived dissociation curves fitting the mean $P_{50}$ (3.68 %) in blue and the ±2σ of $P_{50}$ in red and black (3.34 % and 4.08 % respectively). In each case, the fit has been performed by changing the dissociation constant and keeping the Hill coefficient constant. The $P_{50}$ values reported in Fig. 5c show a good agreement to the ones reported in literature (mean $P_{50}$ for a normal adult individual is 3.7 % [2]). The distribution is somewhat asymmetric, having a larger tail at low $P_{50}$. These high $P_{50}$ cells have a greater oxygen affinity and will bind to oxygen in the lungs more efficiently, but will be more reluctant to release oxygen to the tissues. Statistical analysis of the p50 distribution and its correlation with HbC can be found in *S2*.

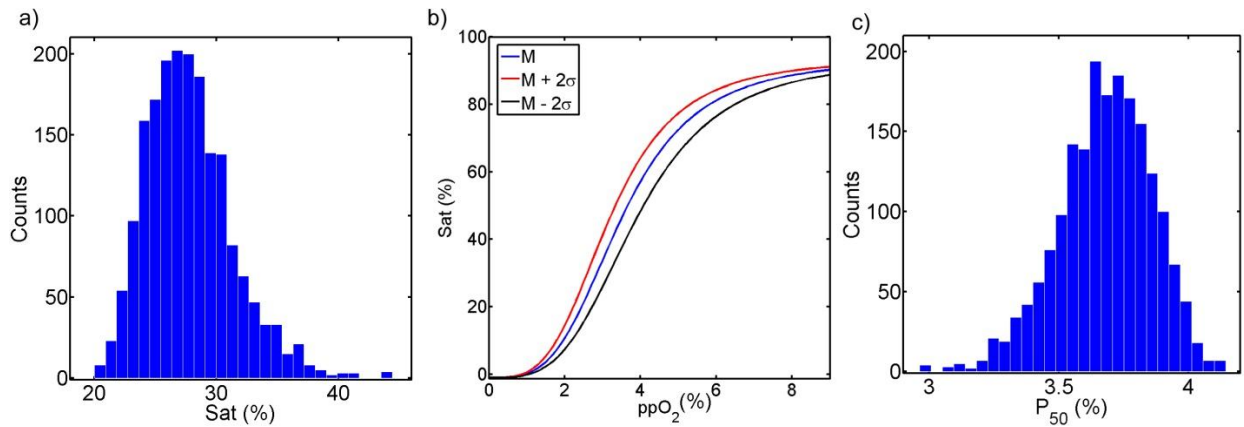

Fig. 5 Single-cell $P_{50}$ distribution. a) Shows the saturation distribution at $PO_2$ = 2.92 %, saturation mean (M) is 27.8 % and σ is 3 %. b) Shows M (in blue), and the values M ± 2σ (in red and black respectively) at $PO_2$ = 2.92 %. c) Shows the corresponding $P_{50}$ distribution where the mean $P_{50}$ is 3.68 %.

Discussion

We have developed an optical measurement system that quantifies volume, hemoglobin mass and oxygen saturation for individual red blood cells in high throughput, while simultaneously accurately controlling oxygen partial pressure. Delivering oxygen is the primary function of a red blood cell and we believe this measurement could constitute a third red blood cell index in clinical hematology analysis, along with single cell volume and hemoglobin concentration, see *S3* for correlations of all three red blood cell indices.

Using this system, we have observed measurable differences in oxygen affinity on a cellular level. We have found that single-cell saturation is positively correlated with single-cell hemoglobin concentration, but that oxygen affinity is not perturbed by changing the osmotic pressure. Instead of an intrinsic relationship between oxygen affinity and hemoglobin concentration, we believe that variability in single-cell oxygen saturation is due to intercellular differences in the 2-3 DPG concentration. Oxygen affinity is modulated by the ratio of the number of 2-3 DPG molecules to the number of hemoglobin molecules in each cell, which does not change under osmotic perturbation.

It is possible that some amount of variability in cellular oxygen affinity could be advantageous. There is a large gradient of ppO$_2$ across the body and a distribution of cellular oxygen affinity could enable a more uniform delivery of oxygen. In addition, under hypoxic conditions, it may be valuable to have a subpopulation of cells with high oxygen affinity to transport oxygen at low ppO$_2$. It is also possible that too much variability is detrimental and could be used as a medical diagnostic for hematological disorders. Similarly, measurement of the P$_{50}$ distribution for a cell population could yield valuable insights into the distribution of hemoglobin variants that occur in various forms of anemia [29].

While we have primarily focused in this paper on taking measurements of cells after they have reached equilibrium with their environment, the methods developed in this study will also be valuable for studying the binding kinetics of single cells before equilibrium has been reached. This could enable a better understanding of the roll of fluid dynamics in oxygen transport and the extent to which single cell parameters affect these kinetics [30]. It will also be valuable to quantify oxygen affinity under different temperatures and pH in order to better understand how these factors effect single cell oxygen affinity and its regulation.

Experimental Procedures

**Device fabrication.** The devices are fabricated using two separate masters, where each is lithographically patterned into SU8 on a silicon substrate. The mold with the blood flow channel is spin-coated with a 100 μm layer of PDMS pre-polymer (1:10 mixture of base and curing agent of Sylgard 184 by Dow Corning, spun at 1000 rpm for 40 s) and partially cured for 3 hrs at 70 °C. The gas layer mold is used to cast 4 mm thick chips of PDMS (1:10 mixture of Sylgard 184, cured overnight at 60 °C). After the inlet and outlet holes were punched in this latter chip, it is aligned and bonded by oxygen-plasma treatment to the flow layer mold. The two-layer device is then separated from the master and the flow layer port holes are punched. To complete the microfluidic devices, the chips are bonded to a microscope cover glass by oxygen-plasma treatment.

**Sample Preparation.** Whole blood from healthy donors was obtained from an outside supplier (Research Blood Components, Brighton, MA). Blood was collected in an EDTA vacutainer, shipped on ice, then rinsed with phosphate buffer saline (PBS) and spun down at 2,000 RPM for 10 min to remove the plasma. To the remaining cells was added 247.5 mL of 40% BSA in PBS, 12 mL of 20% AB9 in PBS, 3 mL of water, and 27.6 mL of 10× PBS, to produce a final osmotic pressure of 272 mOsm, unless otherwise noted, and a red blood cell concentration diluted 60× relative to whole blood. Cells were then loaded into microfluidic channels and measured as a function of ppO$_2$.

**Two wavelength spectroscopy.** To differentiate between HbO$_2$ and Hb, we illuminate with two blue LEDs (L1 λ = 410 nm and L2 λ = 430 nm, bandwidth FWHM = 20 nm). The absorption spectrum of HbO$_2$ and Hb is shown in Fig. 6a [19]. From the absorption spectrum of each species and the power spectral density of each blue LED, we can evaluate the absolute mass of both states of hemoglobin. The values of the Hb mass for the two species are calculated by solving the system of equations:

$$E_{410} = e_{410}^o M_o + e_{410}^d M_d \qquad \text{1(a)}$$

$$E_{430} = e^o_{430}M_o + e^d_{430}M_d \qquad 1(b)$$

where Mo and Md are the masses of oxygenated and deoxygenated hemoglobin, and $e^{o/d}_{410/430}$ is the molar extinction coefficient for HbO$_2$/Hb at 410/430 nm. The region in which the cell is present is segmented using an image processing routine based on an automatic Otsu's threshold evaluation. $E_{410/430} = A_{410/430} \cdot pa \cdot mw$, where A is the Hb absorbance at the two wavelengths defined as $ln\left(\frac{I_{cell}}{I_{ref}}\right)$, where $I_{cell}$ is the segmented image of the cell and $I_{ref}$ is an image when there is not a cell in the field of view, pa is the area of the segmented region containing the cells, and mw is the molecular weight of Hb [18].

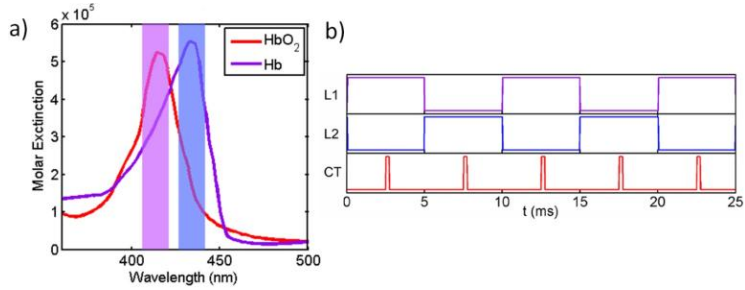

Fig. 6 Two wavelength spectroscopy. a) Absorption spectrum of HbO$_2$ and Hb, where the two vertical bands represent the emission bandwidth of L1 and L2 ($\lambda$ = 410 nm and L2 $\lambda$ = 430 nm respectively, bandwidth FWHM = 20 nm). b) Electronic signals generated by an Arduino microcontroller driving the two LEDs (L1 and L2) and the camera trigger (CT).

We note that we observe un-calibrated saturation values that range from approximately 80% at 21% ppO$_2$ to 0% at 0% at ppO$_2$. We believe that the saturation at 21% ppO$_2$ should be 100% and that our measurement is slightly biased by a small amount of residual optical scattering. In order to conform to this expectation, we have multiplied every cell saturation value by 1.28. We also note that although the measured absorbance values are always positive, the resulting Mo values can be less than zero at 0% ppO$_2$ due to a small amount of measurement noise.

Cells are imaged through a microscope objective (40×, Olympus UPlanFLN, NA 0.75) onto a color camera (AVT Pike, 640×480, pixel size 7.4 µm). The Bayer filter itself allows the independent acquisition of the red and the blue signals. To separate the information carried by the absorption of hemoglobin at 410 and 430 nm, the two LEDs are alternately triggered by an electronic signal (frequency = 100 Hz) generated by an Arduino microcontroller (L1 and L2, Fig. 6b). The same circuit generates the trigger signal for the camera (CT) at double the frequency, enabling the capture of each color image in consecutive frames. The system allows a throughput of approximately 1000 cells per minute.

Supporting information

Cells with high Hb concentration (HbC) show high saturation values. The measurement has been repeated on several different samples from different individuals and each displays a positive correlation between HbC and cell saturation at pressures that correspond to the peak in variability. We report in Fig. S1 the scatter plots of saturation vs. HbC measured on 4 different individuals at $PO_2$ = 2.9 %.

We report in Fig. S2 a statistical analysis on the $P_{50}$ distribution shown in Fig. 6. The dotted line represents a linear fit of the correlation between $P_{50}$ and HbC. In the legend we report the slope of the fitting line, the coefficient of variation of HbC and $P_{50}$, as well as the correlation coefficient of these two variables.

Complete blood counts commonly show a scatter plot of red blood cell volume and hemoglobin concentration, enabling their correlation to be displayed graphically. Figure S3a shows volume and concentration scatter plots collected by our system. The measurements show a negative correlation between volume and hemoglobin concentration, which has previously been studied. We introduce $P_{50}$ as a further RBCs index, and we show in Fig. S3b-d how it correlates with hemoglobin concentration, cell mass and volume respectively.

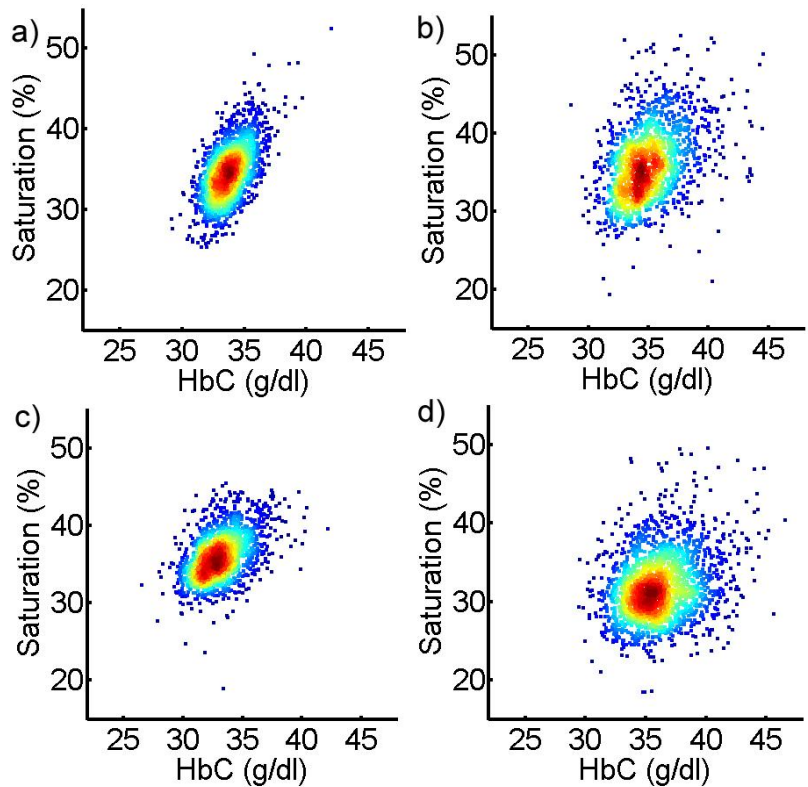

Fig S1. Comparison of saturation distribution at $PO_2$ = 2.9 % among different individuals.

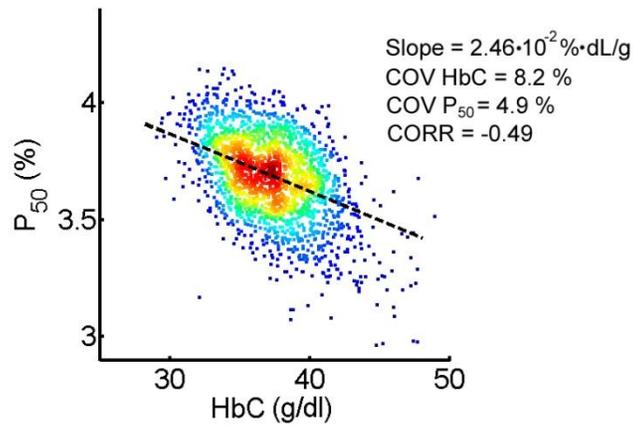

Fig S2. Statistical analysis of single cell $P_{50}$ vs HbC.

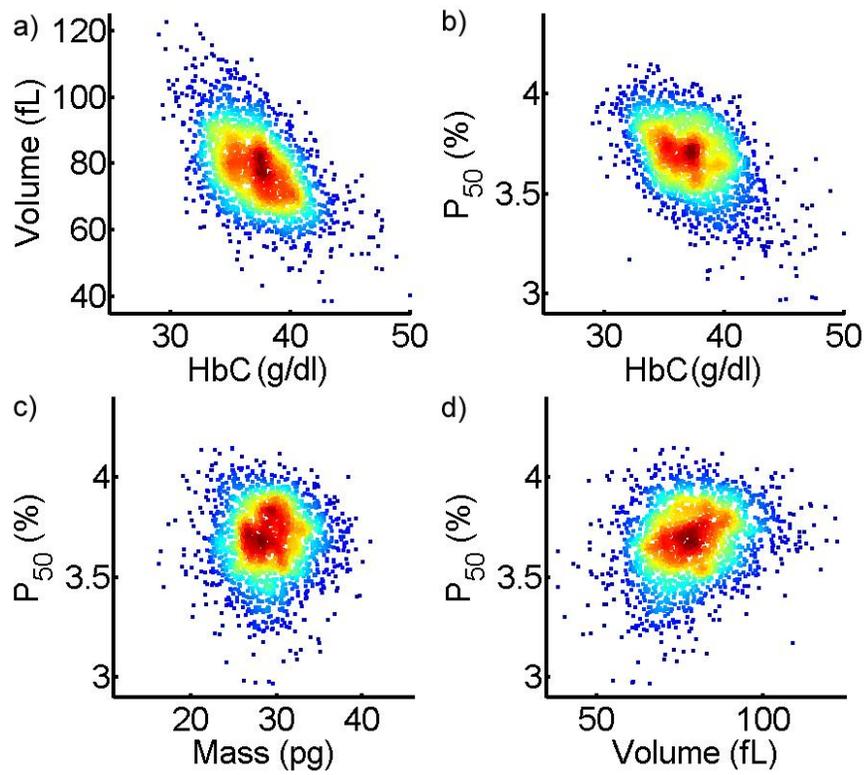

Fig. S3. $P_{50}$ as a third RBCs index.